\documentclass{aa}

\usepackage{graphicx}
\usepackage{txfonts}
\usepackage{xcolor}
\usepackage{hyperref}

\begin{document}

   \title{Co-spatial velocity and magnetic swirls in the simulated solar photosphere}


   \author{Jiajia Liu
          \inst{1}
          \and
          Mats Carlsson
          \inst{2, 3}
          \and
          Chris J Nelson
          \inst{4}
          \and
          Robert Erd{\'e}lyi
          \inst{1, 5, 6}
          }

   \institute{Solar Physics and Space Plasma Research Centre (SP2RC), School of Mathematics and Statistics, The University of Sheffield, Sheffield, S3 7RH, UK\\
              \email{jj.liu@sheffield.ac.uk}
         \and
             Rosseland Centre for Solar Physics, University of Oslo, P.O. Box 1029 Blindern, NO-0315 Oslo, Norway
         \and 
            Institute of Theoretical Astrophysics, University of Oslo, P.O. Box 1029 Blindern, NO-0315 Oslo, Norway
        \and
            Astrophysics Research Centre (ARC), School of Mathematics and Physics, Queen's University, Belfast, BT7 1NN, Northern Ireland, UK
        \and
            Department of Astronomy, E\"{o}tv\"{o}s Lor\'{a}nd University, Budapest, P\'{a}zm\'{a}ny P. s\'{e}t\'{a}ny 1/A, H-1117, Hungary
        \and
            CAS Key Laboratory of Geospace Environment, Department of Geophysics and Planetary Sciences, University of Science and Technology of China, Hefei, Anhui 230026, China
             }

   \date{Received ** **, 2019; accepted ** **, 2019}

 
  \abstract
   {Velocity or intensity swirls have now been shown  to be widely present throughout the photosphere and chromosphere. It has been suggested that these events could contribute to the heating of the upper solar atmosphere, via exciting Alfv{\'e}n pulses, which could carry significant amounts of energy. However, the conjectured necessary physical conditions for their excitation, that the magnetic field rotates co-spatially and co-temporally with the velocity field, has not been
verified.}
   {We aim to understand whether photospheric velocity swirls exist co-spatially and co-temporally with photospheric magnetic swirls, in order to demonstrate the link between swirls and pulses.}
   {The automated swirl detection algorithm (ASDA) is applied to the photospheric horizontal velocity and vertical magnetic fields obtained from a series of realistic numerical simulations using the radiative magnetohydrodynamics (RMHD) code Bifrost. The spatial relationship between the detected velocity and magnetic swirls is further investigated via a well-defined correlation index (CI) study.}
   {On average, there are $\sim$63 short-lived  photospheric velocity swirls (with lifetimes mostly less than 20 s, and average radius of $\sim$ 37 km and rotating speeds of $\sim$2.5 km s$^{-1}$) detected in a field of view (FOV) of 6$\times$6 Mm$^{-2}$, implying a total population of velocity swirls of $\sim$1.06$\times$10$^7$ in the solar photosphere. More than 80\% of the detected velocity swirls are found to be accompanied by local magnetic concentrations in intergranular lanes. On average, $\sim$71\% of the detected velocity swirls have been found to co-exist with photospheric magnetic swirls with the same rotating direction.}
   {The co-temporal and co-spatial rotation in the photospheric velocity and magnetic fields provide evidence that the conjectured condition for the excitation of Alfv{\'e}n pulses by photospheric swirls is fulfilled.}


   \maketitle
%

\section{Introduction}
Swirling motions have been widely observed at different heights in the solar atmosphere, from the photosphere up to the corona \citep[e.g.][]{Wang1995, Bonet2008, Attie2009, Wedemeyer2009, Bonet2010, Wedemeyer2012, Su2014, Dominguez2016, Kato2017, LiuJ2019_ASDA, LiuJ2019_NAT, Shetye2019}. Considering their ubiquity in the solar atmosphere, a number of attempts have been made to explore their potential as an energy supplier to the upper solar atmosphere. Analytical analysis and numerical simulations have suggested that upwardly propagating magnetohydrodynamic (MHD) waves (including magneto-acoustic, e.g. sausage and kink, and especially Alfv{\'e}n waves) could be excited by photospheric swirls \citep[e.g.][]{Carlsson2009, Jess2009, Fedun2011, Shelyag2013, Shukla2013, Mumford2015, Leonard2018, Murawski2018}. These waves may channel significant energy flux into the upper solar atmosphere. Moreover, it has been suggested that swirling motions could also lead to mass flows \citep[e.g.][]{Kitiashvili2012} and result in ubiquitous small-scale solar eruptions \citep[i.e. spicules; e.g.][]{Kitiashvili2013}. In most of the above studies, local magnetic concentrations at the locations of swirls are mandatory, either for the excitation of MHD waves or for the triggering of mass eruptions via mechanisms that employ the Lorentz force.

\begin{figure*}[tbh!]
\centering
\includegraphics[width=0.8\hsize]{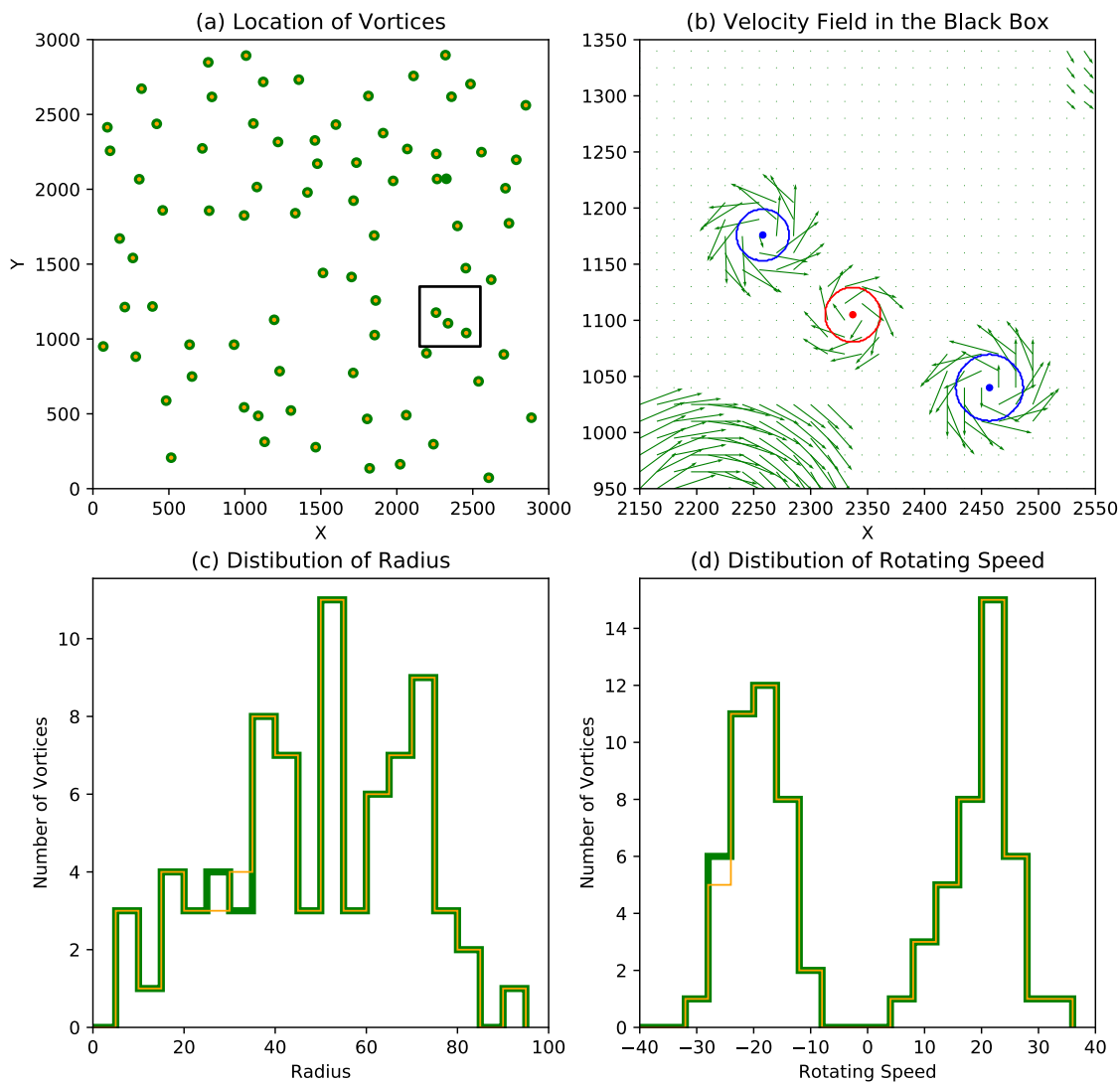}
\caption{Statistics of the synthetic data and the corresponding vortex detection results by ASDA. Green and yellow dots in panel (a) are the locations of  generated and detected vortices, respectively. Panel (b) shows a close-up view of the black box in panel (a). Green arrows represent the velocity field. Blue (red) dots and curves are the centres and edges of vortices with anticlockwise (clockwise) rotations. Panels (c) and (d) are the distributions of the radii and rotating speeds of the generated (green) and detected (yellow) vortices.}
\label{fig_synthetic}
\end{figure*}

Small-scale magnetic concentrations, especially magnetic bright points (MBPs) with magnetic field strengths up to kilogauss (kG) levels, have also been observed to be ubiquitous in the solar photosphere \citep[e.g.][]{Dunn1973, Solanki1993, deWijn2009, Utz2017, LiuYX2018, Keys2019}. Statistical studies have shown that under the current limit of resolution of observations, an MBP has an average radius of the order of 100 km \citep[e.g.][]{Almeida2004, Crockett2010, LiuYX2018} and an average horizontal velocity of the order of 1 km s$^{-1}$ \citep[e.g.][]{Keys2011, Chitta2012}.

\cite{LiuJ2019_ASDA} have recently developed an automated swirl detection algorithm (ASDA, with source code available for free download\footnote{\href{https://github.com/PyDL/ASDA}{https://github.com/PyDL/ASDA}} and applied it to the photospheric observations from both the Solar Optical Telescope \citep[SOT,][]{Tsuneta2008} on board Hinode \citep{Kosugi2007}, and the CRisp Imaging SpectroPolarimeter \citep[CRISP,][]{Scharmer2006} at the Swedish 1-m Solar Telescope \citep[SST,][]{Scharmer2003}. Applying ASDA to photospheric observations sampled by SOT with a pixel size of $\sim$39.2 km, has suggested a total number of 1.62$\times$10$^5$ intensity swirls (essentially velocity swirls, but with the velocity field estimated from the temporal variation of intensities) in the photosphere. Their average rotating speed \citep[obtained by Fourier local correlation tracking, FLCT;][]{Welsch2004, Fisher2008} and radius were estimated to be $\sim$0.9 km s$^{-1}$ and $\sim$300 km, respectively. More than 70\% of the detected intensity swirls were found to be located in intergranular lanes. The similarities between the location, ubiquity, size, and horizontal velocity of MBPs and photospheric intensity swirls promisingly suggest that photospheric swirls might indeed correlate to local magnetic concentrations.

In order to study the propagation of swirling motions from the photosphere to chromosphere using  high-resolution observations, \cite{LiuJ2019_NAT} further applied ASDA to simultaneous photospheric and chromospheric intensity images. A new approach for estimating the correlation between photospheric and chromospheric intensity swirls was developed. A peak correlation was found between photospheric and chromospheric intensity swirls with a time lag of $\sim$130 s. Furthermore, associated numerical simulations capturing the basic key properties of the lower solar atmosphere confirmed that ubiquitous Alfv{\'e}n pulses could be excited by photospheric velocity swirls in a self-similar expanding magnetic flux tube. It was found that these Alfv{\'e}n pulses need approximately 120 s to travel from the photosphere to the upper chromosphere, where similar intensity swirls can be detected. Additionally, these Alfv{\'e}n pulses were found to carry more than enough energy to heat the local upper chromosphere, although no dissipation mechanism was included in the model. It is also worth noting that these numerical simulations required a strong magnetic field concentration at the footpoint region of the magnetic flux tube in order for the Alfv{\'e}n pulses to propagate upward to the upper chromosphere;  without a strong, compact magnetic field the wave energy was reflected back to the photosphere. The next step in accessing this Alfv{\'e}n pulse hypothesis, therefore, is to provide evidence of 1) the co-existence of photospheric velocity swirls and small-scale magnetic concentrations in the solar photosphere and 2) co-temporal swirling motions in the small magnetic concentrations.

Previously, the term magnetic swirl or vortex was used to refer to intensity swirls that co-exist with photospheric magnetic concentrations \citep[e.g.][]{Shelyag2011, Requerey2018}. However, little evidence was shown of the actual swirling motion of the magnetic concentrations themselves. Thus, if the conjecture is verified, we suggest it more appropriate to call these intensity swirls reported in the literature  `magnetized swirls'. In this paper, we define the term magnetic swirl as a region where the magnetic field itself displays a swirling or vortex pattern.

In this work, we apply ASDA to the data from a set of realistic numerical simulations to detect photospheric velocity and magnetic swirls and to analyse their spatial correlations. The paper is organized as follows. The numerical simulation is introduced in Section~\ref{num}. The ASDA code is briefly presented in Section~\ref{met} with further tests on synthetic data. The results are presented in Section~\ref{res}. The conclusion and discussion of results are given in Section~\ref{con}. 

\section{Numerical simulation} \label{num}
Data used in this study are from a series of realistic numerical simulations made using the Bifrost \citep{Gudiksen2011} code. Bifrost was developed with the capability of solving the MHD equations in three-dimensional space, taking into account the effect introduced by radiative transfer in the energy balance equation. The calculations are performed in parallel on a staggered grid using a fifth- or sixth-order compact finite difference scheme. Bifrost has been widely applied to simulate a variety of events in the solar atmosphere \citep[e.g.][]{Carlsson2016}.

The simulation analysed here is 6~Mm $\times$ 6~Mm horizontally with 3~km cell size, and extends vertically from 2.5~Mm below the average height where $\tau_{500}=1$ (defined as $z=0$) to 620~km above this height with 6~km spacing. The horizontal boundary conditions are periodic and the vertical boundaries are transparent with the average pressure tending to a constant value at the bottom boundary. The radiation is treated with 24 rays with the multi-group opacity method of \citet{1982A&A...107....1N} with four opacity bins, extended to treat scattering following \citet{2000ApJ...536..465S}. Background opacities are given by the Uppsala opacity package \citep{Gustafsson1973}. To achieve a relaxed state close to the bottom boundary, the simulation was run for 9.6 hours solar time at lower resolution and then for 8250~s at 3~km horizontal resolution before we started our analysis. The initial magnetic field was a seed field of 0.1~G which was increased through local dynamo action to saturation with an average unsigned magnetic flux of 60~G in the photosphere at $z=0$. In the analysis we included 440 snapshots at 10~s cadence, of which the first frame is labelled  frame 0.

In this study, we  applied ASDA to the horizontal velocity field ($v_x$ and $v_y$) at the average height where $\tau_{500} = $ 1 to search for and detect velocity swirls present in the photosphere. Considering that in practice line-of-sight (LOS) magnetic field observations are much easier to obtain than vector magnetic field observations, we  used only the vertical magnetic field ($B_z$) from the simulations at the average height where $\tau_{500} = $ 1 to detect magnetic swirls.

\begin{figure}[t!]
\centering
\includegraphics[width=0.8\hsize]{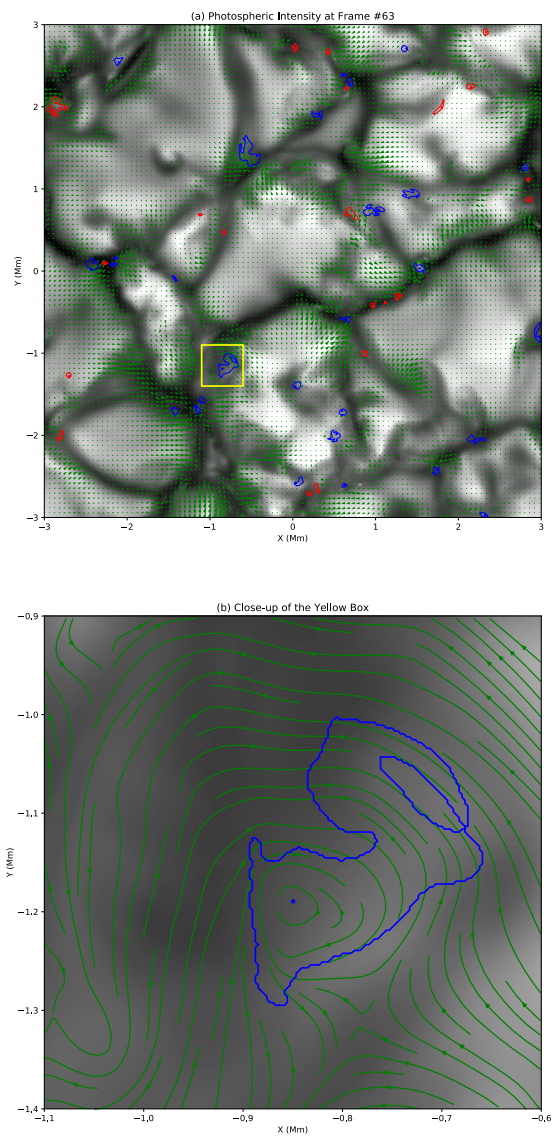}
\caption{Detected velocity swirls at frame 63. The black-and-white background in panel (a) represents the photospheric intensity. Green arrows indicate the photospheric horizontal velocity field from the numerical simulation. Blue (red) dots and curves are the centres and edges of the detected velocity swirls with anticlockwise (clockwise) rotations. Panel (b) is a close-up view of the yellow box in panel (a).}
\label{fig_swirl_example}
\end{figure}

\section{Swirl detection method and tests} \label{met}

The  ASDA developed by \cite{LiuJ2019_ASDA} is used on the photospheric data of the above numerical simulation to detect swirls. ASDA consists of two basic steps to perform swirl detection on a scientifically ready dataset: 1) estimating the horizontal velocity field using the Fourier Local Correlation Tracking (FLCT) method \citep{Welsch2004, Fisher2008}; and 2) applying vortex identification \citep{Graftieaux2001} to the velocity field. If the horizontal velocity field is given in either observations or simulations, the first step can be  skipped. Here, we briefly describe the vortex identification algorithm proposed in \cite{Graftieaux2001} and implemented in \cite{LiuJ2019_ASDA}.

For each pixel $P$ in the image, \cite{Graftieaux2001} proposed two dimensionless parameters:

\begin{equation}
\begin{array}{l}
\displaystyle \Gamma_1(P) = \mathbf{\hat{z}} \cdot \frac{1}{N} \sum_S{\frac{\mathbf{n}_{PM} \times \mathbf{v}_M}{|\mathbf{v}_M|}},  \\
\displaystyle \Gamma_2(P) = \mathbf{\hat{z}} \cdot  \frac{1}{N} \sum_S{\frac{\mathbf{n}_{PM} \times (\mathbf{v}_M - \mathbf{\overline{v}})}{|\mathbf{v}_M  - \mathbf{\overline{v}}|}}.
\end{array}
\label{eq_gamma}
\end{equation}

\noindent Here  $S$ is a two-dimensional region with $N$ pixels containing the target point $P$, $M$ is a point within the region $S$, $\mathbf{\hat{z}}$ is the unit normal vector perpendicular to the observational surface pointing towards the observer, $\mathbf{n}_{PM}$ is the unit radius vector pointing from point $P$ to $M$, $\mathbf{\overline{v}}$ is the average velocity vector within the region $S$, and $\mathbf{v}_M$ is the velocity vector at point $M$. The symbols $\times$ and $|\ |$ are the cross product and the module of vectors, respectively. It was shown that $|\Gamma_2|$ is larger than $2/\pi$ within the edge and $|\Gamma_1|$ peaks at the centre of a swirl \citep{Graftieaux2001}. The sign of $\Gamma_2$ (the same as the sign of $\Gamma_1$) defines the rotating direction of the swirl, with a positive (negative) $\Gamma_2$ and $\Gamma_1$ for  anticlockwise (clockwise) rotation. To find all swirls in a given velocity field, \cite{LiuJ2019_ASDA} proposed to contour the levels of $\pm2/\pi$ of $\Gamma_2$ to find  all the candidates of the swirls. Then the candidates with peak $|\Gamma_1|$ greater than  0.89 are confirmed as swirls. We note that the given threshold only keeps candidate swirls whose   expanding or shrinking speeds are smaller than half of the rotating speeds \citep{LiuJ2019_ASDA}.

In \cite{LiuJ2019_ASDA}, the region $S$ was selected as a square with a size of 7$\times$7 px$^2$. A series of tests were carried out on a number of synthetic datasets containing 1000 randomly generated vortices with various levels of background noise. It was shown by the tests that ASDA may detect fewer swirls than actually present in the data, but   with almost no false detections and  with a high accuracy in determining the parameters of the swirls. However, we  also note that vortices generated in these tests have an average radius of about 7.2 px. Considering that the pixel size of the numerical simulation used in the present work is less than 1/10 of that of the observational data from SOT used in \cite{LiuJ2019_ASDA}, a natural question  arises: Will ASDA still be accurate when the average size of swirls (in units of pixels)  is significantly larger?

Figure~\ref{fig_synthetic}a shows the location (green dots) of 80 vortices in a region with a size of 3000$\times$3000 px$^2$. The radii and rotating speeds of these vortices obey Gaussian distributions, with means and standard deviations of 50 px and 20 px, and 20 px per frame and 5 px per frame, respectively. All these vortices have been put into the image with random positions. Yellow dots in Fig.~\ref{fig_synthetic}a denote the location of vortices detected by ASDA, revealing a detection rate of $\sim$98.8\%, with only 1 out of 80 vortices missing. The green arrows in Fig.~\ref{fig_synthetic}b are the velocity fields of the black box in Fig.~\ref{fig_synthetic}a. Red (blue) curves and dots are the edges and centres of negative (positive) vortices detected by ASDA. It is clear that the edges and centres of the detected vortices both match the original ones well. Also shown,  in Fig.~\ref{fig_synthetic}c and d, are the distributions of the radii and rotating speeds of all generated  and  detected vortices. The location accuracy, radius accuracy, and rotating speed accuracy of the detection \citep[defined by Eq. 5 in][]{LiuJ2019_ASDA} is 100\%, 99.5\%, and 100\%, respectively.

The above results suggest that we can still expect  good performance of ASDA when swirls with various sizes (radius from several to 100 px) exist in the observations.

\section{Results} \label{res}

\begin{figure}[t!]
\centering
\includegraphics[width=0.8\hsize]{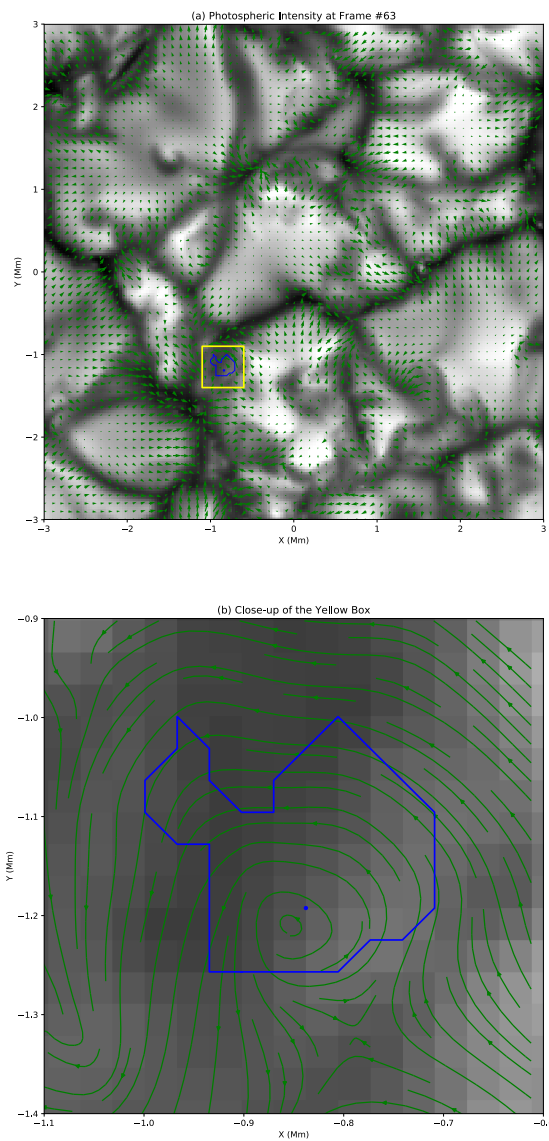}
\caption{Same as Fig.~\ref{fig_swirl_example}, but with the pixel size downgraded to $\sim$32 km.}
\label{fig_swirl_downgrade}
\end{figure}

\subsection{Photospheric velocity swirls}

Figure~\ref{fig_swirl_example}a shows an example of photospheric velocity swirls detected from the photospheric horizontal velocity field in the simulation at frame 63;  Fig.~\ref{fig_swirl_example}b gives  a close-up view of the swirl in the yellow box. The black-and-white   backgrounds in the panels are the corresponding photospheric intensities; also shown are the horizontal velocity fields, and  the edges and centres of negative (positive) swirls. In total, 50 swirls have been detected in this frame, among which around 42\% (21) rotate in the negative direction. Moreover, most swirls are  located in intergranular lanes.

\cite{LiuJ2019_ASDA} found a number density of $3.21\times10^{-2}$ Mm$^{-2}$ of swirls in the photosphere from the Bifrost simulation, which had a pixel size of $\sim$31.25 km. This means that in an image with a field of view (FOV) of 6$\times$6 Mm$^2$, like the one shown in Fig.~\ref{fig_swirl_example}a, we can expect that only one swirl will be detected when the pixel size is downgraded to 31.25 km. Figure~\ref{fig_swirl_downgrade} shows the swirl detection results from the same photospheric horizontal velocity field at frame 63, but with the pixel size enlarged  11 times (downgraded to $\sim$32 km). Unsurprisingly, only one swirl (surrounded by the yellow box) was detected from the downgraded data.

\begin{figure*}[tbh]
\centering
\includegraphics[width=\hsize]{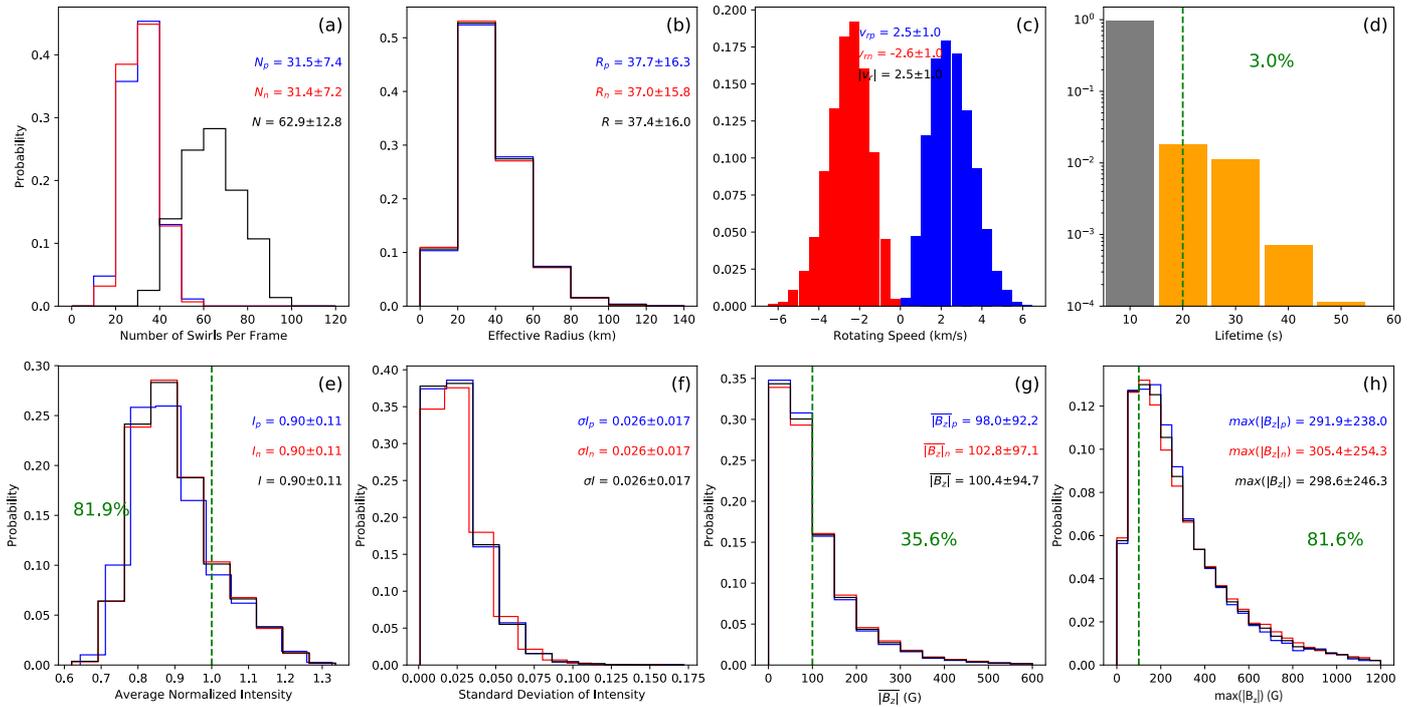}
\caption{Statistics of the photospheric velocity swirls detected by ASDA from the numerical simulations. In panels (a), (b), (c), (e), (f), (g), and (h) swirls with negative (positive) rotations are shown in red (blue), and all swirls are shown in black. $N$ denotes the number of swirls in each frame of the simulation, $R$ the average effective radius, $v_r$ the average rotating speed, $I$ the average normalized photospheric intensity within the area of each swirl, $\sigma I$ the standard deviation of the normalized photospheric intensity of each swirl, $\overline{|B_z|}$ the average absolute vertical magnetic field strength of each swirl, and $max(|B_z|)$ the maximum absolute vertical magnetic field strength of each swirl. The green vertical lines in panels (d), (e), and (g)--(h) depict a lifetime of 20 s, an average normalized photospheric intensity of 1, and a vertical magnetic field strength of 100 G, respectively.}
\label{fig_stat_vel}
\end{figure*}

In total, 27627 photospheric velocity swirls were detected from 439 frames of photospheric horizontal velocity maps,   about 49.9\%  of which (13781 swirls) rotate in the negative direction. We note that in order to perform correlation analysis between the detected velocity swirls and magnetic swirls (Section~\ref{sec_correlation}), swirl detection was not performed on the last frame (440) of the velocity maps. The above result means that  on average there are 62.9$\pm$12.8 swirls in one frame (Fig.~\ref{fig_stat_vel}a) with a FOV of 36 Mm$^2$, corresponding to a number density of 1.75$\pm$0.36 Mm$^{-2}$ of photospheric velocity swirls. Though the spatial resolution of the numerical simulations used in this study is $\sim$13 times higher than that of the Hinode/SOT observations (39.2 km) used in \cite{LiuJ2019_ASDA}, the number density is about 65 times higher. This indicates an estimated total number of 1.06$\times$10$^7$ of velocity swirls in the photosphere.  Figure~\ref{fig_stat_vel}b depicts the distribution of effective radii of all photospheric velocity swirls detected, showing negative, positive, and all swirls. Knowing the edge and centre, the effective radius of a swirl is defined as the radius of the circle that has the same area as the swirl \citep[Eq. 6 in][]{LiuJ2019_ASDA}. The average effective radius of all swirls is around 37.4$\pm$16.0 km, which is only 1/8th of that of the Hinode/SOT photospheric intensity swirls \citep{LiuJ2019_ASDA}. The above results again confirm the conclusion made in \cite{LiuJ2019_ASDA} that spatial resolution has a vital influence on the number and size of the detected swirls.

The  distribution of the rotating speeds of all detected swirls is shown in Fig.~\ref{fig_stat_vel}c. The distributions of positive and negative swirls are both Gaussian, with an average absolute rotating speed of $\sim$2.5 km s$^{-1}$. This value is about 2.8 times  that of the photospheric intensity swirls studied in \cite{LiuJ2019_ASDA}. We note that the photospheric horizontal velocity fields in \cite{LiuJ2019_ASDA} were determined from the FLCT, and it was suggested that local correlation tracking could underestimate the photospheric horizontal velocity field by a factor of as much as 3 \citep{Verma2013}.

We employed the method proposed in \cite{LiuJ2019_ASDA, LiuJ2019_NAT} to estimate the lifetime of swirls at a single height. For two swirls, S$_1$ detected at time $t_0$ and S$_3$ at time $t_0+2\Delta t$, where $\Delta t$ is the cadence of the observation, S$_1$ and S$_3$ are considered to be the same swirl if the expected location of the centre of S$_1$ after $2\Delta t$ is located within S$_3$:

\begin{equation}
\mathbf{c_1} + 2\mathbf{v_{c1}}\cdot \Delta t \subset S_3.
\label{eq_life}
\end{equation}

\noindent Here, $\mathbf{c_1}$ and $\mathbf{v_{c1}}$ are the location and velocity of the centre of swirl S$_1$, respectively. The above criterion is applied repeatedly to the swirl detection results from the first frame to the last  to evaluate the lifetimes of all swirls detected. The distribution of lifetimes of the detected photospheric velocity swirls is shown in Fig.~\ref{fig_stat_vel}d. Because $\sim$97\% of the swirls have only been  detected in one frame with lifetimes of less than 20 s, we were unable to conclude on a reliable average lifetime for them under the limit of the current cadence. The result that most swirls have lifetimes of less than 20 s is consistent with what has been found in \cite{LiuJ2019_ASDA}. We  note that the lifetimes of swirls discussed in the present work are estimated at a single height ($\tau_{500}=1$). Considering that swirling motions usually propagate upwards \citep{LiuJ2019_NAT}, the timescale of swirling motions travelling from the photosphere to the chromosphere is on average $\sim$ 130 s, which is much longer.

The photospheric intensity of the simulation is the intensity of the continuum bin in the multi-group opacity method \citep{1982A&A...107....1N}; therefore, there is no meaningful physical unit for the photospheric intensity obtained from the simulation. In the statistics, the normalized intensity is used,  calculated  by dividing the intensity by the average photospheric intensity of all frames (0.228). The average normalized photospheric intensity of the detected velocity swirls has a mean value of around 0.90$\pm$0.11 (Fig.~\ref{fig_stat_vel}e), and the standard deviation of the normalized intensity of swirls has a mean value of around 0.026$\pm$0.017 (Fig.~\ref{fig_stat_vel}f). The standard deviation of the photospheric intensity of a detected swirl is then, on average, less than 3\% of its average photospheric intensity, meaning that the intensities of these swirls are homogeneous in the first-order approximation. This may lead to fewer swirls being detected when applying local correlation tracking methods to intensity observations to estimate the horizontal velocity field. For a detected swirl, we determine that it is located in the intergranular lanes if its average normalized photospheric intensity is less than one (green dashed line in Fig.~\ref{fig_stat_vel}e). We find that at least 82\% of the detected swirls are located in intergranular lanes.

The mean value of the average absolute photospheric vertical magnetic field strength of all detected photospheric velocity swirls is around 100 G. $\sim$36\% swirls have average absolute vertical magnetic field strengths above 100 G (Fig.~\ref{fig_stat_vel}g), indicating that the detected velocity swirls very likely correspond to local magnetic concentrations. This can be seen more clearly in Fig.~\ref{fig_stat_vel}h from the distribution of the maximum absolute photospheric vertical magnetic field strengths of the detected photospheric velocity swirls. The mean value of the maximum absolute vertical magnetic field is around 300 G, with $\sim$82\% swirls having maximum absolute vertical magnetic field strengths above 100 G. More intriguingly, of the swirls with   maximum absolute vertical magnetic field strengths above 100 G, more than 85\% have average normalized photospheric intensity values of less than one. These results suggest that most swirls are spatially correlated to local magnetic field concentrations in intergranular lanes with magnetic field strengths above 100 G.

The question we ask now is the following: Are these magnetic field concentrations magnetic swirls?

\subsection{Correlation between velocity swirls and magnetic swirls} \label{sec_correlation}

\begin{figure*}[tbh]
\centering
\includegraphics[width=\hsize]{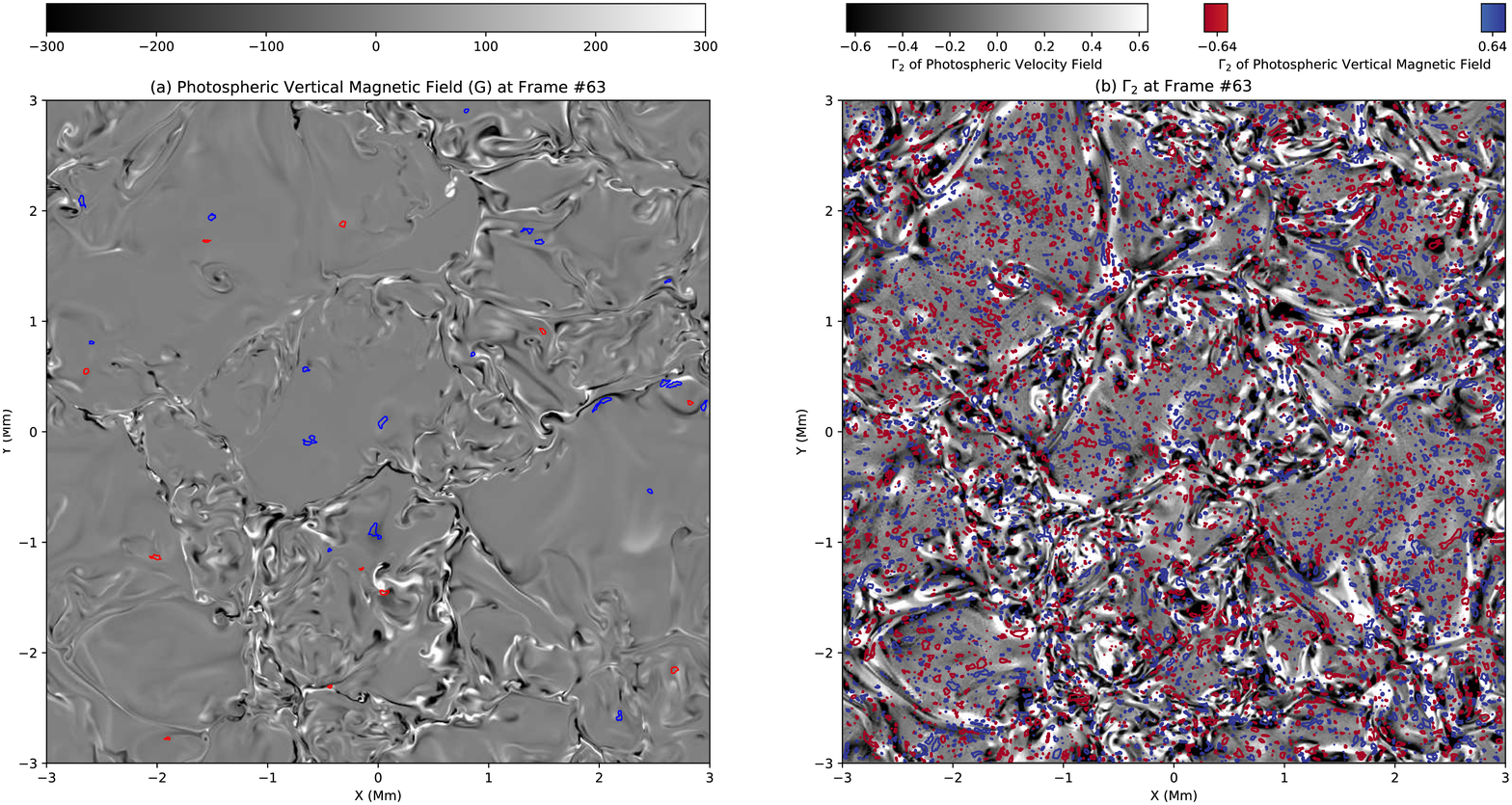}
\caption{Example of the detected magnetic swirls. The black-and-white background in panel (a) is the photospheric vertical magnetic field at frame 63. Blue (red) dots and curves are the centres and edges of the detected magnetic swirls with a anticlockwise (clockwise) rotation. The black-and-white  background in panel (b) denotes the $\Gamma_2$ distribution obtained from the photospheric horizontal velocity field at frame 63. The red and blue contours in panel (b) represent the $\Gamma_2$ distribution obtained from the photospheric vertical magnetic field at levels of $-2/\pi$ and $2/\pi$, respectively.}
\label{fig_bz_example}
\end{figure*}

To detect photospheric magnetic swirls from the simulations, we first need  to determine the velocity field of the movement of photospheric magnetic elements. FLCT version 1.06 (\href{http://cgem.ssl.berkeley.edu/cgi-bin/cgem/FLCT/index}{http://cgem.ssl.berkeley.edu/cgi-bin/cgem/FLCT/index}) is applied to the photospheric vertical magnetic field to achieve this goal, with the experimental `bias correction' algorithm turned on. The bias correction algorithm was introduced into FLCT version 1.06 as an effort to correct the artificially low amplitudes of the horizontal velocity field returned by FLCT. Preliminary tests on comparisons between the velocity fields obtained from the photospheric intensity by FLTC and directly from the numerical simulation suggests that turning on the bias correction can increase the calculated horizontal velocity field strength by a factor of around 80\%. However it still underestimates the real velocity field by a factor of $\sim$3. We note that the two scalars $\Gamma_1$ and $\Gamma_2$ (see Eq.~\ref{eq_gamma}) used to find swirls do not dependent on the actual values of the horizontal velocity field strength.

The  black-and-white  background in Fig.~\ref{fig_bz_example}a shows the photospheric vertical magnetic field at frame 63.  A comparison between Fig.~\ref{fig_bz_example}a and Fig.~\ref{fig_swirl_example}a clearly reveals that most strong magnetic field concentrations are located within intergranular lanes, just as expected. Again, red (blue) curves and dots denote the edges and centres of the detected magnetic swirls with negative (positive) rotations. Clearly, most magnetic swirls are found located in strong local magnetic concentrations.

It was demonstrated in \cite{LiuJ2019_NAT} that there are a number of difficulties in directly comparing swirls detected at two different layers $L_1$ and $L_2$ (here, $L_1$ is the photospheric horizontal velocity field and $L_2$ the photospheric vertical magnetic field) to find where they overlap with each other. These difficulties include, but are not limited to, the following: 1) swirls confirmed by ASDA are only a part of all the candidates, as some candidates have been removed due to large expanding or shrinking speeds, and 2) the shapes of swirls are usually irregular, suggesting that a velocity swirl and its corresponding magnetic swirl (if present) are unlikely to be 100\% overlapped, even if they are at exactly the same location.

\begin{figure*}[tbh!]
\centering
\includegraphics[width=\hsize]{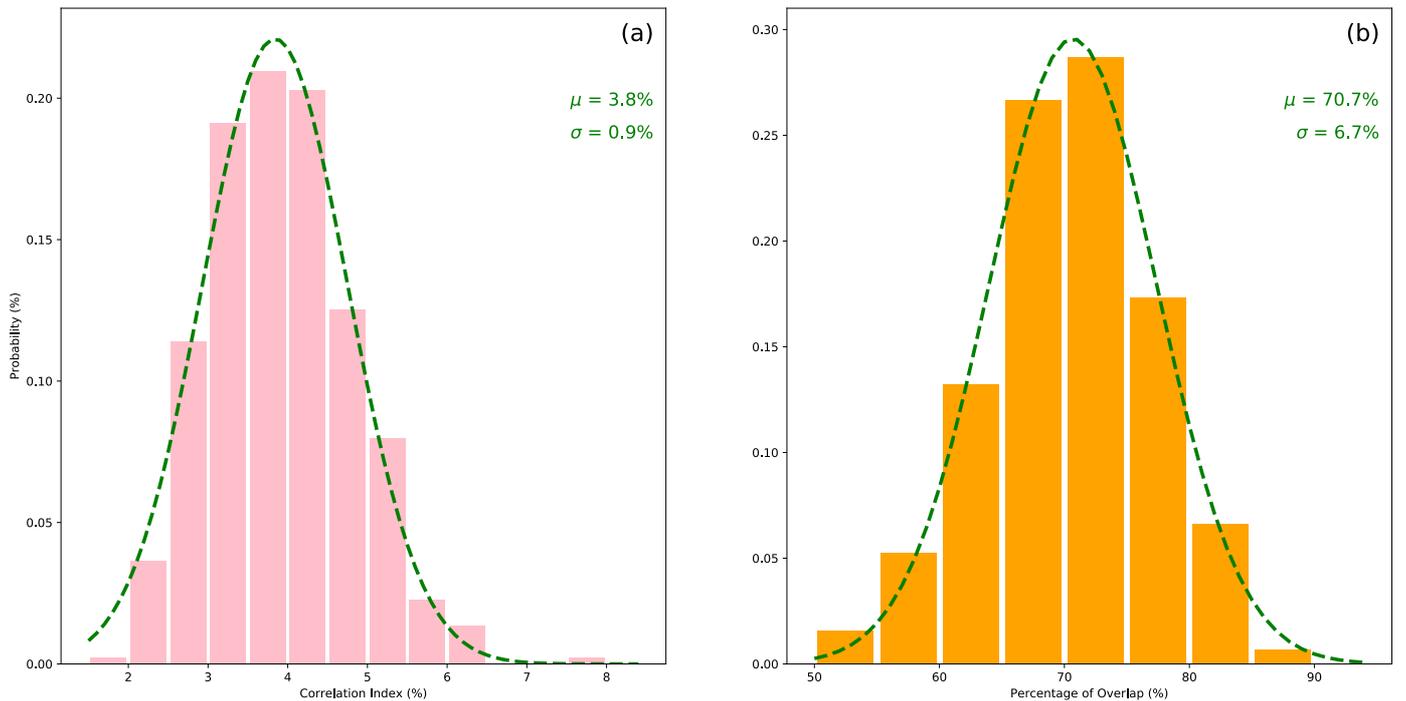}
\caption{Correlation between photospheric velocity and magnetic swirls. Panel (a) denotes the distribution of the correlation index (CI) between $\Gamma_2$ maps from co-temporal photospheric horizontal velocity and vertical magnetic field maps. Panel (b) is the distribution of the percentage of velocity swirls in each frame, which overlap with photospheric magnetic swirls. The green dashed lines in both panels are the corresponding Gaussian fit results, and $\mu$ and $\gamma$ are the mean and standard deviation of the Gaussian fit, respectively.}
\label{fig_correlation}
\end{figure*}

Here we employ the method proposed by \cite{LiuJ2019_NAT}, which was used successfully to find the correlations between photospheric and chromospheric intensity swirls in order to study the correlations between the velocity and magnetic swirls. The method is briefly summarized   here:

\begin{itemize}
    \item  All points that are greater (less) than $2/\pi$ ($-2/\pi$) in the $\Gamma_2$ maps of $L_1$ and $L_2$ are set to be 1 (-1). All other points are set to 0. The resulting $\Gamma_2$ maps of layer $L_1$ and $L_2$ are named  $\Gamma_2^1$ and $\Gamma_2^2$, respectively. Figure~\ref{fig_bz_example}b shows the $\Gamma_2$ maps obtained from the photospheric velocity field (black-and-white background) and the photospheric vertical magnetic field (red-blue contours);\\
    
    \item   $T_1$ is defined as the sum of the absolute values of all points in $\Gamma_2^1$, and $T_2$ as that in $\Gamma_2^2$;  $T$ is set as the minimum of $T_1$ and $T_2$;\\
    
    \item The correlation map $C$ is obtained by multiplying $\Gamma_2^1$ and $\Gamma_2^2$ point by point. The correlation index ($CI$) between layer $L_1$ and $L_2$ is then defined as $(\sum{C}) / T$. The definition determines that $CI$ is always within the range between -1 and 1, and a higher $CI$ suggests a higher correlation;\\
    
    \item For a swirl $S$ detected in layer $L_1$, its points are mapped onto the $C$ map to calculate the percentage of points that have positive $C$ values. The percentage is then defined as the correlation index ($CI_S$) of swirl $S$. Swirl $S$ is marked to be overlapped with a swirl candidate in layer $L_2$, if $CI > 0$ and $CI_S > t_h$, where $CI$ is defined as in the previous step. Here, $t_h$ is defined as $\Delta^2/\overline{A}$, where $\overline{A}$ is the average area of the detected swirls ($37.4^2\pi$ km$^2$ $\approx 4394$ km$^2$) and $\Delta$ is the pixel size ($\sim$2.93 km). Then, $t_h$ is calculated to be 0.002. The above procedure means that if, on average, there is at least one point within swirl $S$ corresponding to a positive value in the $C$ map, it is then considered to be overlapped with a swirl candidate in layer $L_2$. This process also determines that swirl $S$ in layer $L_1$ and the corresponding swirl in layer $L_2$ must rotate in the same direction.
\end{itemize}

Repeating the above procedure at each time step of the simulation between the $\Gamma_2$ maps of the simultaneous photospheric velocity field and vertical magnetic field, we were able to study the correlation indices ($CI$s) and overlaps between the photospheric velocity and magnetic swirls. For a comparison, we  also calculated the average $CI$ and percentage of velocity swirls in each frame that have co-spatial magnetic swirls, after randomly shuffling the original datasets for ten times, as errors of the measurement. The average $CI$ of the randomly shuffled datasets is $0.09\% \pm 0.02\%$, meaning that any $CI$ less than $\sim$0.19\% is within the 5$\sigma$ range. The average percentage of swirls in each frame that correspond to magnetic swirls of the randomly shuffled datasets is $26.9\% \pm 0.9\%$, meaning that any percentage of overlap less than $\sim$31.4\% is within the 5$\sigma$ range.

Figure~\ref{fig_correlation}a depicts the distribution of the $CI$ between each pair of $\Gamma_2$ maps calculated from simultaneous photospheric velocity and vertical magnetic field maps. The distribution of the $CI$ resembles a Gaussian distribution (green dashed curve), with a mean and standard deviation of about 3.8\% and 0.9\%, respectively. The lowest $CI$ is about 1.8\%, which is well above the upper value of its corresponding 5$\sigma$ range (0.19\%). 

Figure~\ref{fig_correlation}b shows the distribution of the percentage of velocity swirls in each frame that are overlapped with magnetic swirls. Again, the distribution of the percentage of overlap resembles a Gaussian distribution (green dashed curve), with a mean and standard deviation of about 70.7\% and 6.7\%, respectively. The lower limit of the percentage of overlap is $\sim$52.1\%, again, still   well above the upper limit of its corresponding 5$\sigma$ range (31.4\%).

\section{Conclusions and discussion} \label{con}
In this paper, we have applied the automated swirl detection algorithm (ASDA) to photospheric velocity and magnetic fields obtained from a series of high-resolution  Bifrost numerical simulations (with a pixel size of $\sim$2.93 km). In this section we  now present our conclusions on the major findings and present further discussion.

In total, 27627 velocity swirls, with approximately half rotating in the positive direction and the other half in the negative direction, have been detected from 439 frames of photospheric horizontal velocity maps, resulting in approximately 63 swirls per frame with a FOV of 6$\times$6 Mm$^2$. This means that there is a velocity swirl number density of 1.75$\pm$0.36 Mm$^{-2}$ in the photosphere, which is much higher than the previously reported swirl number density \citep{LiuJ2019_ASDA}, determined by Hinode/SOT observations with a pixel size   more than ten times larger (39.2 km). We then expect a total number of 1.06$\times$10$^7$ swirls at any given time in the solar photosphere. However, this number needs to be further confirmed by actual observations of the solar photosphere, by an instrument with comparable resolution, such as the Daniel K. Inouye Solar Telescope (DKIST), which will be available from 2020. The average radius of these velocity swirls is estimated to be $\sim$37.4 km, less than the pixel size of the observations with the highest resolutions currently available, namely  from  the Hinode/SOT and the Swedish 1-m Solar Telescope (SST). The average rotating speed of the detected photospheric velocity swirls is $\sim$2.5$\pm$1.0 km s$^{-1}$, which is about three times  that obtained from Hinode/SOT and SST observations, confirming that FLCT usually underestimates the horizontal velocity field.

At least 82\% of the detected photospheric velocity swirls have been found located in intergranular lanes, suggesting again that most photospheric velocity swirls  originated from intergranular lanes. Intriguingly, $\sim$82\% of the detected photospheric velocity swirls are found to have peak absolute vertical magnetic field strengths greater than 100 G, indicating that most photospheric velocity swirls co-spatially exist with strong local magnetic field concentrations.

Applying ASDA to the photospheric vertical magnetic field, evidence of photospheric magnetic swirls have been revealed. Further analysis of the correlation between the detected photospheric velocity and magnetic swirls show that, on average, about 71\%$\pm$7\% of the photospheric velocity swirls have corresponding magnetic swirls. This suggests that most photospheric velocity swirls exist co-spatially and co-temporally with photospheric magnetic swirls. The method used here also determined that a pair of co-existing velocity swirl and magnetic swirl must rotate in the same direction, which is consistent with the frozen-in conditions expected in the solar photosphere.

\cite{LiuJ2019_NAT} presented evidence of the propagation of Alfv{\'e}n pulses excited by photospheric intensity swirls, through studying the correlations between photospheric and chromospheric intensity swirls with the aid of realistic numerical simulations. However, because of the lack of high-resolution photospheric magnetic field observations, evidence of the excitation of Alfv{\'e}n pulses by photospheric intensity swirls were not given. Results in this paper show that most swirls (at least 70\%)  are indeed accompanied by disturbances in their co-spatial local magnetic concentrations. A necessary condition for Alfv{\'e}n pulses to be excited is proved to be fulfilled by the analysis in this paper.

\begin{acknowledgements}
      J.L., C.J.N. and R.E. acknowledge the support from the Science and Technology Facilities Council (STFC, grant numbers ST/M000826/1, ST/L006316/1). C.J.N. also acknowledges support received from the STFC with grant number ST/P000304/1. This research was supported by the Research Council of Norway through its Centres of Excellence scheme, project number 262622 (Rosseland Centre for Solar Physics), and through grants of computing time from the Programme for Supercomputing. J.L acknowledges support from the International Rosseland Visitor Programme. R.E. also acknowledges the support from the Chinese Academy of Sciences President’s International Fellowship Initiative (PIFI, grant number 2019VMA0052).

\end{acknowledgements}

\bibliographystyle{aa}

\end{document}